# Interaction between graphene and copper substrate: The role of lattice orientation

*Otakar Frank[1,*], Jana Vejpravova[2], Vaclav Holy[3], Ladislav Kavan[1], and Martin Kalbac[1]*

[1] J. Heyrovsky Institute of Physical Chemistry of the AS CR, v.v.i., Dolejskova 2155/3, CZ-182 23 Prague 8, Czech Republic

[2] Institute of Physics of the AS CR, v.v.i., Na Slovance 1999/2, CZ-182 21 Prague 8, Czech Republic

[3] Faculty of Mathematics and Physics, Charles University in Prague, Ke Karlovu 5, CZ-121 16 Prague 2, Czech Republic

**Abstract**

We present a comprehensive study of graphene grown by chemical vapor deposition on copper single crystals with exposed (100), (110) and (111) faces. Direct examination of the as-grown graphene by Raman spectroscopy using a range of visible excitation energies and microRaman mapping shows distinct strain and doping levels for individual Cu surfaces. Comparison of results from Raman mapping with X-ray diffraction techniques and Atomic Force Microscopy shows it is neither the crystal quality nor the surface topography responsible for the specific strain and doping values, but it is the Cu lattice orientation itself. We also report an exceptionally narrow Raman 2D band width caused by the interaction between graphene and metallic substrate.

[*] Corresponding author. Tel. +420 266 053 446. email: otakar.frank@jh-inst.cas.cz (Otakar Frank)

The appearance of this extremely narrow 2D band with full-width-at-half maximum (FWHM) as low as 16 cm$^{-1}$ is correlated with flat and undoped regions on the Cu(100) and (110) surfaces. The generally compressed (~ 0.3% of strain) and n-doped (Fermi level shift of ~250 meV) graphene on Cu(111) shows the 2D band FWHM minimum of ~20 cm$^{-1}$. In contrast, graphene grown on Cu foil under the same conditions reflects the heterogeneity of the polycrystalline surface and its 2D band is accordingly broader with FWHM > 24 cm$^{-1}$.

## 1. Introduction

Graphene [1], a monoatomically thin membrane of sp$^2$-hybridized covalently bonded carbon atoms, attracts enormous research efforts due to its unique physical properties, such as high carrier mobility, superior thermal conductivity, high optical transparency and extreme mechanical properties [2-5]. Many applications exploiting these properties have been envisaged or already tested ranging from mechanical resonators [6], integrated circuits [7], solar cells [8, 9], transparent touch displays [10] to terahertz detection [11]. However, prospective applications of such devices require large and high quality graphene layers. In this regard, chemical vapor deposition (CVD) on metals is the most promising way of large-scale production of continuous graphene layers [12-14]. Even though general principles like the role of carbon solubility in the catalyst, pressure or temperature in the CVD chamber are established [15-18], a detailed understanding of the interplay between the surface structure of the metal and the resulting graphene is still lacking. Copper, in the form of a thin polycrystalline foil, is the most widely employed catalyst due to its low cost and ease-of-use for obtaining graphene monolayers with reasonable quality, with single-crystalline graphene domains in sizes of several millimetres [19]. The influence of the crystalline nature of Cu foils on the domain sizes and orientation, layer number and quality of the grown graphene has been shown recently [20-24]. To avoid the presence of Cu grain boundaries and transition between individual Cu lattice orientations, both of which largely affect the resulting graphene, epitaxial growth on surfaces with defined long-range crystalline order is particularly appealing [15, 18, 25-32]. The studies conducted on single crystals may provide direct information about the effects of Cu lattice on the graphene growth as a function of pressure/temperature, without the perturbations from Cu grain boundaries.

Raman spectroscopy is undoubtedly one of the prominent methods used for characterization of graphene [33, 34]. It can distinguish between single-layer, bilayer and multilayer graphene, it is highly sensitive to doping [35-37] and mechanical deformation [38-41]. Moreover, the capabilities of Raman spectroscopy of CVD graphene are vastly expanded by using precursors with varying content of $^{12}$C and $^{13}$C isotopes [42, 43]. The usual way of Raman characterization of CVD graphene transferred to Si/SiO$_2$ suffers from the transfer process (ripples or even fissures, contamination, etc.), and also the substrate can substantially modify both the electronic and phonon structure of the overlaying graphene. Hence the knowledge of Raman characteristics of as-grown graphene is of utmost importance not only for a viable *ex post* monitoring of the growth process but also from the fundamental point of view regarding the relationship between substrate and graphene. With a few exceptions [20, 27, 44, 45], most of the Raman spectra are measured on graphene transferred to Si/SiO$_2$ substrate instead of directly on Cu, due to a much larger Raman scattering intensity and lower spectral background on Si/SiO$_2$. The differences in graphene growth conditions (including substrate orientation), are in turn smeared and not resolved.

In this study we focused on graphene grown on Cu(100), (110) and (111) single crystals by a low pressure CVD from methane with either $^{12}$C or $^{13}$C isotope. Large-area microRaman mapping together with a thorough statistical evaluation of the data (G and 2D band shifts, linewidths and intensities, 2D band dispersion) are compared against the individual surfaces and their qualities and orientations obtained by X-ray diffraction (Laue diffraction method, real and reciprocal space mapping) and AFM topography. All three Cu surface orientations show a different impact on the overlying graphene concerning the amount and uniformity of strain and doping. However, the most notable result consists in a remarkable and unprecedented narrowing of the 2D band, especially for the graphene grown on Cu(100). The possible reasons are uniformity of the strain/doping fields and interaction between graphene and the substrate.

## 2. Experimental

### 2.1 Sample preparation

The graphene samples were synthesized using CVD as reported previously [43]. In brief: The Cu single crystals (MaTecK, purity 99.9999%, orientation accuracy < 0.1°) were heated to 1000 °C and annealed for 20 min under flowing $H_2$ (50 standard cubic centimeter per minute (sccm)). Then the crystals were exposed either to 1 sccm $^{12}CH_4$ or $^{13}CH_4$ for 20 minutes leaving hydrogen gas on with the same flow rate of 50 sccm. Finally the substrate was cooled down still under $H_2$ flow by opening the furnace and removing the reaction tube from the heated zone (cool down time to 500°C is ~ 5 minutes). The pressure was kept at 0.35 Torr during the whole growth. All three Cu single crystals were subjected to the CVD at the same time, close to each other in the furnace to avoid differences in the growth conditions.

### 2.2 Sample characterization

The copper single-crystals were checked by various X-ray diffraction techniques. First, the crystallographic orientation was verified by the standard Laue method using a CHIRANA – Micrometa device with Cu Kα radiation. The single-crystals were mounted with the (100), (110) and (111) plane, respectively, perpendicular to the beam path and the diffraction pattern was recorded in the back scattering geometry. The beam size was fixed to 3 mm in diameter. Quality of each sample was further investigated by mapping the characteristic diffractions (hkl) of the copper *fcc* structure in the reciprocal space: (400), (200) for the (100) face, (220), (222) for the (110) face, and (111), (222) for the [111] face. The reciprocal maps were recorded on the X'Pert PRO MRD XL in high-resolution set-up, the multiple symmetric 2Theta-Omega scans were recorded. The data were analyzed using Matlab software considering a mosaic block model with a coherent crystal-truncation rod simulated by a two-dimensional PearsonVII function yielding the characteristic vertical and lateral size of the mosaic blocks and their mean misorientation, respectively [46, 47]. The influence of multiple heating cycles (equal to the conditions of the CVD) on the crystallinity was investigated by the reciprocal mapping. In order to probe the grain distribution within the ingot area, the mapping of the intensity of a single characteristic diffraction for each orientation was performed in the real space. First, the orientation of each single-crystal was optimized to maximum intensity of the diffraction at a single point and then the intensity of the diffraction was mapped within the sample area (10 x 10 mm$^2$) with a step of 1 mm in the x and y directions, respectively. AFM images were recorded using a Dimension Icon Microscope (Bruker) with ScanAsyst Air tips in the PeakForce tapping mode.

The Raman spectra were excited by He-Ne (1.96 eV) or Ar$^+$-Kr$^+$ (2.71, 2.54, 2.41, 2.34 or 2.18 eV) lasers and acquired by a LabRam HR spectrometer (Horiba Jobin–Yvon) with a pixel-to-pixel spectral resolution of approximately 1 to 2 cm$^{-1}$ depending on the excitation energy (from 1.96 to 2.71 eV). The spectrometer was interfaced to a microscope (Olympus, 100x objective) and the laser power was kept lower than 1 mW under the objective to avoid heating of the sample. Intensity response of the CCD detector was calibrated by tungsten halogen light source HL-2000-CAL (Ocean Optics) for each excitation. Raman peaks were fitted by Lorentzian line shapes for the analysis with a prior subtraction of a linear background in predefined spectral ranges (1200-1700 cm$^{-1}$ for D and G band fitting, 2400-2900 for 2D band fitting). The accuracy of peak position determination, as acquired by repeated measurements on one point is ~ ±0.6 cm$^{-1}$ for 1.96 eV excitation, which corresponds well to the pixel-to-pixel spectral resolution. The accuracy of FWHM determination was found to be also ±0.6 cm$^{-1}$. Raman mapping was conducted with lateral steps of 1 μm (both in X and Y directions) on rectangular areas with varying sizes - minimum number of 110 data points and 200 data points on average for each map. For details on the map sizes, see Table S1 (Supplementary Information). Accumulation time for each spectral window was 120 s. The samples were measured under ambient conditions, and stored in a vacuum desiccator.

Due the relatively low intensity of Raman features of graphene on Cu, $^{13}$C was used to check whether there are no interfering peaks from the background which might contribute to the analysed bands (D, G, and 2D). No such peaks were observed, apart from the vibration of molecular oxygen at 1554 cm$^{-1}$. The spectra acquired on $^{13}$C graphene were used in the same manner as those from $^{12}$C, only the band positions were recalculated for the sake of direct comparability according to the equation:

$$(\omega_0 - \omega)/\omega_0 = 1 - [(12 + c_0)/(12 + c)]^{1/2} \tag{1}$$

where $\omega_0$ is the frequency of the particular Raman mode in the $^{12}$C sample, $c = 0.99$ is the concentration of $^{13}$C in the enriched sample, and $c_0 = 0.0107$ is the natural abundance of $^{13}$C.

The determination of electron concentration (for electron doping) from the G band position was done by approximating the data from ref. [35] by a quadratic polynomial $\Delta Pos(G) = -0.986n^2 + 9.847n$, where $\Delta Pos(G)$ is the shift from the "strain" line (see Results and Discussion) and $n$ is the electron concentration in $10^{13}$ cm$^{-2}$. The corresponding Fermi level shift was determined as $\Delta E_F = \sqrt{\pi n v_F^2}$, where $v_F$ is the Fermi level velocity.

## 3. Results and Discussion

Graphene samples were synthesized using CVD as reported previously [48] on Cu (100), (110) and (111) single crystals and on a polycrystalline Cu foil for comparison. Cu surface quality was checked by XRD real and reciprocal space mapping, and Laue diffraction method.

The first simple check of the crystal orientation and quality was carried out by the Laue method (typical Laue patterns are shown in Figure S1, Supplementary Information). The patterns reflect the characteristic Laue class symmetry: *m-3m* expected for the (100), (110) and (111) planes in the *fcc* structure (space group no. 225). The reciprocal maps for all orientations around the two selected reflections are presented in Figures S2-S4 (Supplementary Information). The fit to the simple mosaic model is also shown and the resulting values are summarized in Table 1. The characteristic vertical size of the crystal block is identical for all directions within the experimental error, while the lateral size varies between 400 – 800 nm. The mean misorientation angle is *ca.* 0.2°, 0.1° and 0.2° for the (100), (110) and (111) orientations, respectively. Considering a typical size of a graphene grain (20 - 50 μm) [43, 48], the mean block size is at least one order of magnitude smaller. Hence, the fine structure variations of the copper crystal (within a single grain) should significantly influence neither the growth nor the interaction of the graphene with the copper surface.

The latter observation is also supported by the studies of the annealed samples. The reciprocal maps for the (100) and (110) orientation before and after annealing (1000° C, $H_2$) are depicted in Figure S5 (Supplementary Information). For the (100)-oriented single-crystal, the diffuse scattering becomes significantly suppressed after annealing, which suggests improvement of the crystallinity. For the (110)-crystal, the diffuse scattering after annealing is almost negligible, which also points to a substantial improvement of the structure. The observed behavior is in correspondence with the well-known effect of polygonization in copper upon annealing [49].

Table 1: Results of the XRD reciprocal map analysis using the simple mosaic model.

|       | Vertical block size (nm) | Lateral block size (nm) | Misorientation ($^o$) |
|-------|--------------------------|-------------------------|------------------------|
| (100) | 100 ± 20                 | 400 ± 20                | 0.20 ± 0.05            |
| (110) | 100 ± 20                 | 800 ± 20                | 0.10 ± 0.05            |
| (111) | 100 ± 20                 | 600 ± 20                | 0.20 ± 0.05            |

Finally, the distribution of the macroscopic grains within the ingot was analyzed using the real space mapping, the results are shown in Figure S6 (Supplementary Information). Typically, the ingot contains several grains of few mm in diameter with misorientation below $0.4^o$.

Graphene grown on the copper surfaces was studied in detail by microRaman spectroscopy. The samples were measured freshly after graphene growth or stored under vacuum in the meantime. However, to check for the effects of possible oxidation, e.g. during manipulation or measurement, i) we varied the order of measurement of samples in every experimental batch (see Table S1, Supplementary Information), and ii) we controlled the appearance of the samples under optical microscope. As an example, Figure S7 (Supplementary Information) shows the difference between graphene on Cu(111) freshly grown and after one week of storing under ambient conditions. The optical microphotograph of freshly grown graphene is essentially featureless due to high reflectivity of the Cu surface, with the exception of very faint and sparse darker spots belonging to bilayer graphene patches. The area looks similarly after 12 hours of Raman mapping. On the other hand, more features can be observed on the aged sample, where dark spots and thin lines appear due oxidation of the Cu surface, mainly on the defective grain boundaries but with an only slow progression into the grain interiors, in line with the previous observations [50, 51]. Hence, the short (hours) copper oxidation causes only marginal changes of the Raman spectra. However, a partial and local influence cannot be ruled out, as proposed earlier [30].

Figure 1 shows typical Raman spectra of the G and 2D (or G') regions of graphene as-grown on Cu(100) and (111) acquired using laser energies from 1.96 eV (633 nm) to 2.71 eV (457 nm). The measurements were taken on spots, which correspond to the most frequently appearing spectra (in terms of peak positions, widths and frequencies) for the particular Cu surface based on statistical evaluation of Raman maps (see further). The spectra of graphene grown on Cu(110) are generally similar to those on Cu(100) and are shown in Fig. S8 (Supplementary Information). The D band region is not shown due to the absence or very low intensity of the D peak in these spectra. It should be noted that the D band is present occasionally throughout the samples on the graphene grain boundaries or ad-layers [52]. Throughout the following text, the frequencies of G and 2D bands will be referred to as Pos(G) and Pos(2D), respectively, full-widths-at-half-maxima as FWHM(G) and FWHM(2D), and I(G) or I(2D) denote intensities of the bands expressed as integrated areas. Lorentzian lineshapes were used for fitting. If not stated otherwise, the results referring to a certain Cu orientation are meant for graphene measured as-grown on the particular surface.

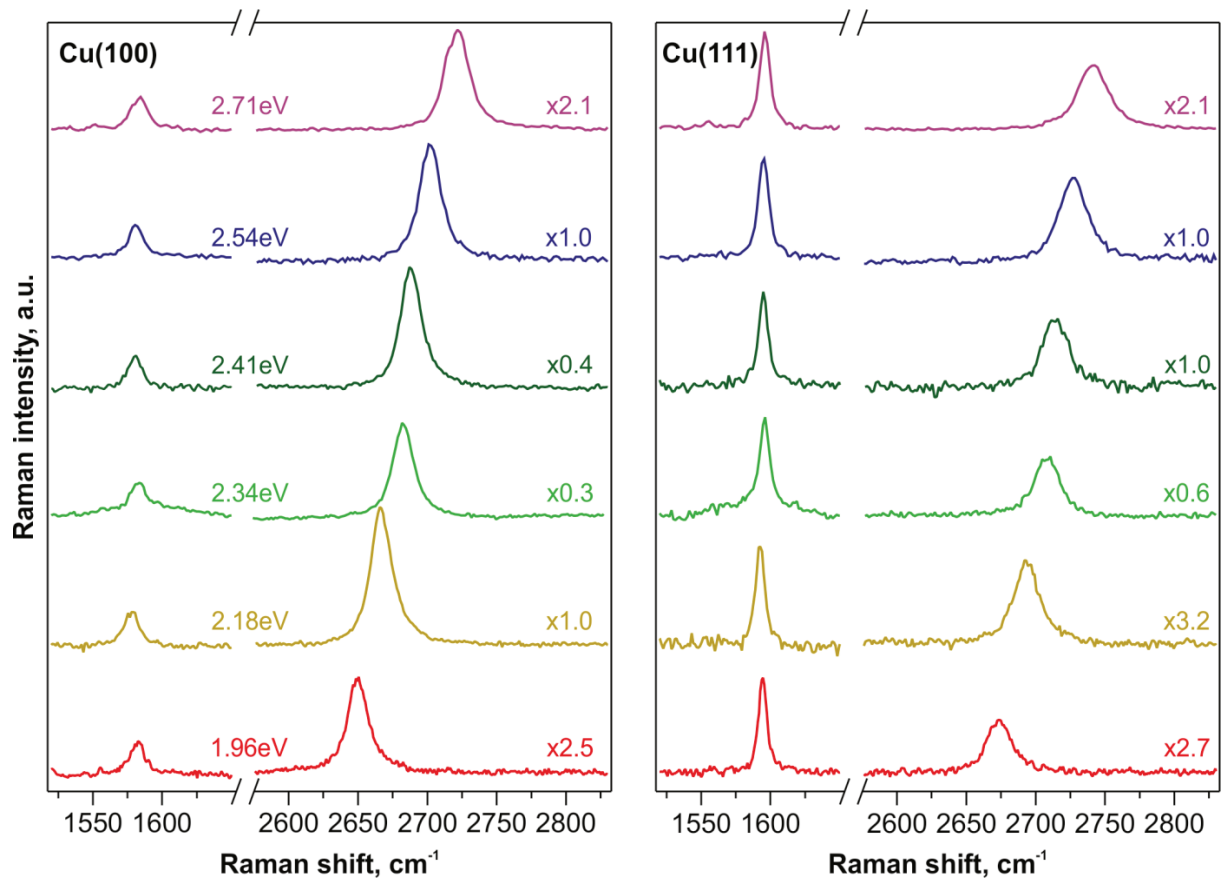

**Figure 1.** Raman spectra of graphene grown on Cu(100) and (111) single crystals. Excitation energies are given for each spectrum, the spectra are normalized to the G band intensity for each Cu surface, the corresponding zoom factors are labeled in the plots.

Fig. 1 shows that there are several differences in the spectra of graphene grown on the two different Cu substrates. The I(2D)/I(G) intensity ratio (evaluated as integrated areas, 4 different measurements for each excitation) is higher for Cu(100) and Cu(110), where it amounts to 5.2 ± 0.7 and 5.0 ± 0.7, respectively, with no statistically significant dependence on the laser excitation energy. The I(2D)/I(G) for graphene on Cu(111) is 2.0 ± 0.4, again without a proportionality to the laser excitation energy. Furthermore, the G band of graphene grown on Cu(111) is upshifted and narrowed compared to that grown on Cu(100), while the 2D band is upshifted but broadened for graphene grown on Cu(111). Selected relations between frequencies, FWHM and $I(2D)/I(G)$ ratios are plotted in Figure 2 for all four Cu surfaces, as acquired from the Raman mapping using the 1.96 eV laser excitation energy. The Raman features of graphene on Cu(100) are grouped similarly to Cu(110) in terms of intensities and frequencies (panels a and b in Fig. 2) and both highly contrast with the features of graphene grown on Cu(111). The results on polycrystalline Cu foil are usually in between the values obtained on Cu(100) and Cu(111) surfaces. The averaged Raman spectra obtained from the mapping are shown in Figure S9 (Supplementary Information).

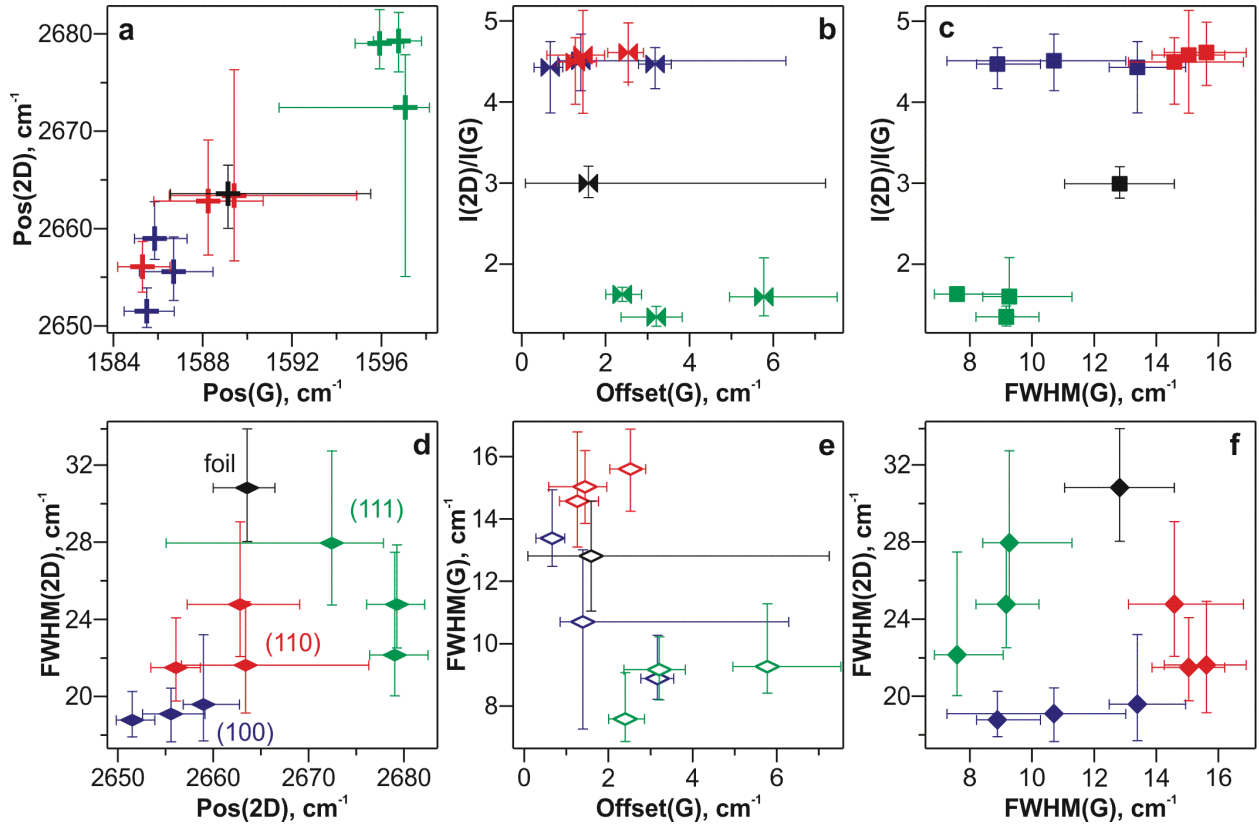

**Figure 2.** Correlations between the monitored parameters of the G and 2D peaks fitted by Lorentzian lineshapes in Raman spectra of as-grown graphene on Cu(100) (blue), Cu(110) (red), Cu(111) (green) and polycrystalline Cu foil (black). Each data point represents a median of all measurements (> 200 map points on average) on a given substrate from one CVD batch. The Offset(G) parameter is derived from Pos(G) by subtracting the G band value on the strain line for a constant Pos(2D), see text. The "error" bars represent the first (Q1) and third quartiles (Q3) of each dataset. Note: Raman shifts from CVD batch using $^{13}CH_4$ are recalculated via Eq. 1 (see Experimental Section) to be comparable with samples produced from $^{12}CH_4$. Laser excitation energy is 1.96 eV.

Besides the extremely narrow FWHM(2D) (panels d and f), which will be discussed later in the text, most of the differences shown in Fig. 2 can be understood in terms of the current theories and previous experiments. The G band originates from a conventional first order Raman scattering process and corresponds to the in-plane, doubly degenerate phonon mode (transverse (TO) and longitudinal (LO) optical) with $E_{2g}$ symmetry at the Brillouin zone center (Γ point) [53]. Its frequency and width are influenced by doping [35, 36] and stress [38], while its integrated intensity was assumed to be dependent only on the excitation energy ($E_L$) as I(G) ∝ $E_L^4$ [54, 55] or I(G) ∝ $E_L^2$ [56]. However, as has been shown recently, quantum interference effects play a crucial role in the G band intensity, where blocking of excitation pathways dramatically increases I(G) under strong doping [36, 57]. The increase in Pos(G) for both electron and hole doped graphene is caused by a non-adiabatic removal of the Kohn anomaly at Γ point [37, 58], and simultaneous FWHM(G) decrease is caused by the Pauli blocking of phonon decay into electron-hole pairs [37]. For suspended graphene, the mean values of Pos(G) = 1580 cm$^{-1}$ and FWHM(G) = 14 cm$^{-1}$ can be regarded as experimental benchmarks for nearly undoped graphene with the estimated upper bound of carrier density $n$ of ~2 × 10$^{11}$ cm$^{-2}$ [59]. Mechanical stress causes G band downshift under tension and upshift under compression with a rate of 57 cm$^{-1}$/% for biaxial strain [60]. Under uniaxial strain the G bands splits into two components G$^-$ and G$^+$ with the shift rates of 31 and 10 cm$^{-1}$/%, respectively [38, 40].

The 2D mode comes from a second-order triple resonant process between non-equivalent K points in the Brillouin zone of graphene, involving two zone-boundary, TO-derived phonons with opposite momenta $q$ and $–q$ [61-63]. Numerous theoretical and experimental works, e.g. [61, 62, 64-68], appeared in the last decade describing various aspects of the origin and properties of this intense, narrow and dispersive line, yet there are still several debatable issues. The 2D band has been regarded as a single Lorentzian with FWHM of ~ 24 cm$^{-1}$ for graphene on Si/SiO$_2$ or broader depending on the substrate and/or environment [34, 69]. As it turned out recently, free standing graphene exhibits narrower (~ 22 cm$^{-1}$) 2D band with a bimodal lineshape [59, 70, 71].

Similar subfeatures were observed also for the unstrained graphene on SU-8 photoresist [72] and in uniaxially strained monolayers [72-74]. In a simplified one-dimensional picture there can be two dominant directions of the phonon wavevectors contributing to the 2D mode – along K-Γ (so called inner) or K-M (outer) high symmetry lines. Earlier works usually assumed a preferential contribution of the latter to the 2D mode [66, 68, 69], but recent works point to a larger importance of the inner phonons [62, 72, 73, 75]. However, the simple one-dimensional picture is now being superseded by a full two-dimensional description of the electronic bands and phonon dispersion and matrix elements [62], in which the notation of inner or outer phonons is of a weaker relevance [67]. The 2D mode is dispersive, and its frequency changes with the excitation energy with the slope $\partial Pos(2D)/\partial E_L$ of approx. 100 cm$^{-1}$/eV [34]. The 2D band is also sensitive to doping and mechanical stress but these effects manifest themselves differently from the case of the G band. Strain causes the 2D band shift in the same directions as the G band, with the shift rates for biaxial strain larger by a factor of ~ 2.2-2.5 [60, 76]. While broadening and splitting accompanies the shift for uniaxial strain [72], such effects are not observed for the biaxial one [60]. Charge injection causes Pos(2D) increase for hole doping with a Pos(2D)/Pos(G) shift ratio of ~0.5-0.7 [35, 36, 77], whereas electron doping causes only a negligible Pos(2D) change for $n \leq 2 \times 10^{13}$ cm$^{-2}$ followed by a non-linear Pos(2D) decrease for higher n-doping levels [35, 36]. The FWHM(2D) increases and I(2D) decreases upon both p- and n-doping mainly due to electron-electron interactions, and also electron-phonon coupling strength (expressed as a dimensionless parameter $\lambda$), which increases to a small extent upon doping [62, 65]. Additionally, the change in $\lambda$ causes the 2D dispersion slope to decrease upon doping [78].

We first consider the Pos(G) in our samples (appearing in Fig. 2a and Figs S10c,d, Supplementary Information). Graphene on Cu(100) exhibits unambiguously the lowest Pos(G), with medians from the three runs between 1586-1587 cm$^{-1}$ with the interquartile range (IQR =

Q3-Q1, see Fig. 2) between 2.2 and 3.3 cm$^{-1}$. Graphene on Cu(110) shows the Pos(G) slightly higher and with a larger variation, medians fall between 1585 and 1590 cm$^{-1}$, with IQR of 3.0 – 8.0 cm$^{-1}$. Pos(G) for graphene on Cu(111) is consistently the highest with medians between 1596 and 1597 cm$^{-1}$ and IQR ~ 2.1 cm$^{-1}$ in the latter two runs. The first run is rather exceptional for Cu(111) (see Fig. 3), when a second group of peaks at substantially lower frequencies (both G and 2D) appeared, occasionally also in the same spectra. The appearance of the lower frequency G and 2D or the simultaneous presence of the two G and 2D bands was spatially random in the maps and their origin is not clear at moment. We assume that these low frequency peaks belong to well defined areas with sizes even below the laser spot, where graphene relaxed from the otherwise compressed state (see below) with a simultaneous formation of wrinkles. The hypothesis is corroborated also by larger FWHMs of both G and 2D bands in the lower frequency group (13.2 vs. 9.5 cm$^{-1}$ and 35.6 vs. 27.2 cm$^{-1}$ for average FWHM(G) and FWHM(2D), respectively), which is another indicator of buckling [38]. The median Pos(G) of graphene grown on Cu foil is 1589 cm$^{-1}$ with a large IQR of 9 cm$^{-1}$, i.e. being the average of the oriented surfaces and spanning their whole range. The correlation between Pos(2D) and Pos(G) is summarized in Fig. 2a and detailed in Fig. 3. The general trends for Pos(2D) are similar to Pos(G): the lowest frequencies are for Cu(100), the highest for Cu(111) with the foil being in between. From plots in Fig. 3, we can see that most of the variations between individual points on each surface are driven by strain – the $\partial$Pos(2D)/$\partial$Pos(G) is consistently ~ 2.5 for the linear correlations [39, 60, 77]. In several cases, the correlation is broadened or even a second set of points with an offset in the Pos(G) direction appears, clearly evidencing additional variations in doping [77]. The relative variation of both the strain and doping levels between the individual surfaces can be estimated from Figs 2 and S11 (Supplementary Information). Considering the "strain" line with slope of 2.5 in the Pos(2D)/Pos(G) plot delimiting the low frequency boundary of data points from Cu(100) and Cu(110), we can estimate the change in charge carrier concentration in our samples. Note that the few datapoints of Cu(100) and Cu(110) with G band frequency downshifted from the "strain" line in Fig. S11 have a large FWHM(2D) and low I(2D)/I(G), hence they probably come from small bilayer patches sometimes appearing in CVD graphene on Cu [48, 79, 80]. For low carrier concentrations, the Pos(2D) change with doping is negligible compared to ∂Pos(G) [35, 36]. The separation between the "strain" line and Pos(G) median for Cu(111) is ~ 3 cm$^{-1}$, corresponding to a change in carrier concentration $n$ of ~ 3 × 10$^{12}$ cm$^{-2}$ or ~ 200 meV shift of the Fermi level ($|\Delta E_F|$), see Experimental section for details of $|\Delta E_F|$ estimation. Using the same method, the two distinct sets of data points within Cu(100) marked by arrows in Fig. S11 exhibit G band upshift from the "strain" line of 4 and 8 cm$^{-1}$, corresponding to $|\Delta E_F|$ of ~ 250 and 350 meV, respectively. We should note that it is impossible to derive the sign of doping at such low levels from this approach due to the aforementioned small ∂Pos(2D). On the other hand, the difference in strain can be easily estimated in our case, as ∂Pos(2D) between Cu(100) and Cu(111) of ~ 23 cm$^{-1}$ gives ~ 0.2% of biaxial compression [38, 40, 60].

The variation in doping is also evidenced in Figs 2b, c and e, which display the mutual dependencies of I(2D)/I(G), FWHM(G) and the G band frequency. Since the role of strain cannot be directly discerned in these plots, we define the Offset(G) as the Raman shift difference between the measured Pos(G) and the "strain" line at the same value of Pos(2D) for the given point. We thus neglect the subtle slope of ∂Pos(2D)/∂Pos(G) caused by doping in the Offset(G) parameter (Fig. 2b and e), because ∂Pos(2D)/∂$n$ is < 1 cm$^{-1}$ per 10$^{13}$ cm$^{-2}$ for electron doping up to 2×10$^{13}$ cm$^{-2}$, hence it is smaller than the measurement accuracy in the encountered range of doping levels. For completeness, Pos(G) instead of Offset(G) is shown in Fig. S10c and d. Graphene on both Cu(100) and Cu(110) has median I(2D)/I(G) close to 4.5 with only a small variation, caused to some extent by the aforementioned distinct doped areas in Cu(100), in contrast to graphene on Cu(111), where the median I(2D)/I(G) is in the range of 1.3-1.7 for the three data sets. The decrease of I(2D) upon doping is in accord with previous experiments [35, 36] and theory [62, 65].

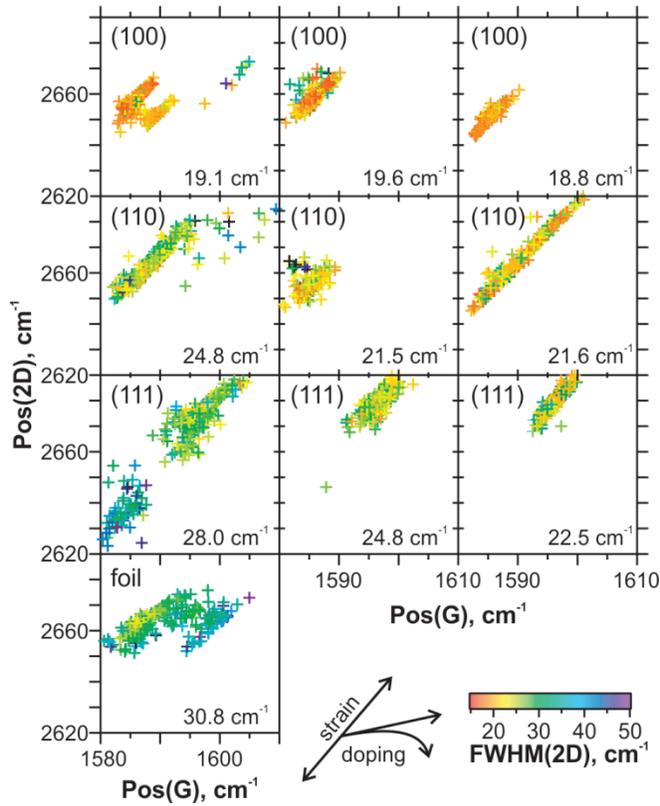

**Figure 3.** Correlation between all measured Pos(2D) vs. Pos(G) for individual surfaces (rows) in each CVD batch (columns). Colors of the data points are scaled according to FWHM(2D) for the particular point with the scale bar in the right bottom corner. The sketch with arrows in the bottom centre depicts the trends for strain and doping, derived from earlier works [35, 36, 38, 40, 60, 77]. The numbers in bottom right corner of each plot represent the median FWHM(2D) for the particular run.

The state of graphene on Cu(100) should be determined from its Pos(G) and Pos(2D), which exhibit the lowest frequencies among the studied samples. Assuming Pos(G) of 1580 cm$^{-1}$ to be the benchmark value for suspended graphene [59, 71], median Pos(G) for Cu(100) is upshifted by ~ 7 cm$^{-1}$. Should the difference be caused solely by strain (~ 0.1% of biaxial compression), the corresponding theoretical ∂Pos(2D) would amount for ~ 18 cm$^{-1}$, which is still even lower than the measured ∂Pos(2D). The higher-than-expected unstrained Pos(2D) can be partially explained by the smaller slope of 2D band frequency dispersion with $E_L$. Note that doping would cause the relative measured ∂Pos(2D) to drop compared to the theoretical "strained" one derived from ∂Pos(G). In addition, the main fraction (~ 60%) of measured points on Cu(100) exhibits FWHM(G) of 13-14 cm$^{-1}$ (Fig. S12, Supplementary Information), hence we might conclude that the majority of graphene on Cu(100) is essentially undoped and slightly compressed (~ 0.1%). Graphene on Cu(110) shows a similar behavior, while graphene on Cu(111) manifests additional 0.2% of compression, i.e. ~ 0.3% in total and >200meV |$E_F$| shift.

Few data are available on the doping level of as-grown graphene on copper. Using angle-resolved photoemission spectroscopy (ARPES), Walter et al. [30] reported on n-doping of graphene on Cu(111) single crystals with $E_F$ shift of ~ -300 meV and additional ~ -300 meV for graphene on Cu(100) presumed to originate from intercalation of oxygen from air [30]. The results closely match our Raman data, *cf.* |$E_F$| ~ 200 meV for Cu(111) and three distinct doping levels in Cu(100) with |$E_F$| ~ 0, 250 and 350 meV. Siegel et al. [81] obtained varying levels of n-doping in graphene grown on Cu foil with the average value of 2 × 10$^{13}$ cm$^{-2}$ corresponding to Δ$E_F$ ~ -500 meV, likewise by means of ARPES [81]. Fermi level shift smaller than 110 meV was found by micro-spot ARPES on graphene grown on (100) facets of a copper foil by Wilson et al. [24]. Using the first-principles density functional theory (DFT), Khomyakov et al. [82] calculated that the differences in work functions between graphene and Cu should cause n-doping in graphene with Δ$E_F$ = -170 meV.

The strain in as-grown graphene on metal substrates is due to the difference between their thermal expansion coefficients. Upon cooling, the metal lattice shrinks, while the graphene lattice expands. Provided there is a good adhesion of graphene on the surface, this inevitably leads to a biaxial compression in graphene. Eventually, the strain can be partially released by the formation of wrinkles, as was documented by STM, e.g. in graphene on Cu(111) [83] or Ir(111) [84]. Another interpretation of strain in graphene on Cu surface is based on the mismatch between the lattice parameter of Cu and equilibrium C-C bond length, when superstructures from graphene at Cu(111) and Cu(100) are formed [26]. Molecular dynamics simulation of the superstructures showed inhomogeneous compressive strain between 0.3 and 0.6% for graphene on Cu(111) and more varying strain for graphene on Cu(100) ranging from 0.3% compression to 0.2% tension [26]. Our data show strain variation of approx. 0.2% for data sets on Cu(100) and Cu(111), and up to ~ 0.4% for Cu(110), see Figs 3 and S11. In one case, the strain on Cu(111) was interpreted as relaxed by wrinkling in parts of the measured area, see above. It should be noted, however, that the molecular dynamics calculated strain [26] is inhomogeneous on an atomic level and thus should be manifested by broader Raman peaks with a narrow frequency distribution. We observe the opposite – narrow Raman peaks but with a wider frequency span, hence documenting strain inhomogeneities on a scale closer to the size of the laser spot (~ 1 µm). On the other hand, given the coincidence between the MD calculated [26] and our measured strain values, the "lattice mismatch" interpretation cannot be ruled out. This could be especially valid for graphene on Cu(111), which shows a non-negligible charge transfer, and we assume the interaction there to be stronger than between graphene and Cu(100) or Cu(110). The small doping level as well as small strain in the latter cases point towards relaxed graphene layers with larger distance from the substrate and hence smaller interaction.

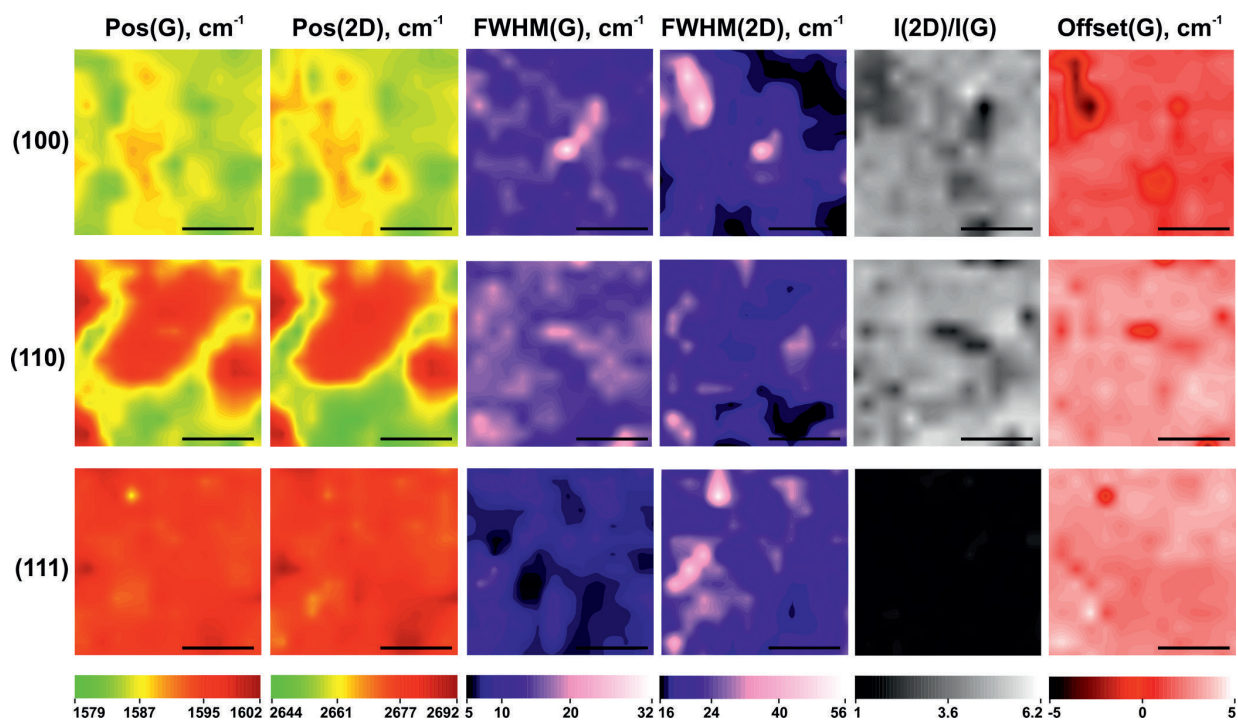

**Figure 4.** Maps parameters obtained from fitted Raman spectra of graphene (3$^{rd}$ batch) grown on Cu(100) – top row, (110) – middle row, and (111) – bottom row. The color scales are always the same for a particular parameter across the Cu faces, as depicted below the respective columns. The dimensions of all the Raman maps are the same, the scale bar is 5 μm. The laser excitation energy was 1.96 eV.

By comparing Raman maps of the monitored G and 2D band parameters (Figure 4) and AFM images taken in the same experimental batch (Figure S13, Supplementary Information), the influence of Cu topography can be discerned in some cases. Firstly, all three copper surfaces exhibit flat regions alternating with steep edges of varying height, originating presumably by evaporation of the metal during the low pressure CVD growth. The spatial extent of these topographic features corresponds well with the spatial distribution of G and 2D band frequencies (2 leftmost columns in Fig. 4). The height of these features derived from AFM profiles (Fig. S13) is then directly reflected in the span of G and 2D band frequencies for the particular surface – the smallest height as well as Pos(G), Pos(2D) differences are shown for Cu(100) and Cu(111), whereas the height profile of Cu(110) shows roughness one order of magnitude larger with the Pos(G) and Pos(2D) spanning over 20 and 50 cm$^{-1}$, respectively. However, no other Raman feature seems to be influenced by the microscale roughness of the Cu crystal. Furthermore, there is no correlation between neither nanoscale nor microscale roughness of the Cu surface and the Raman spectra of grown graphene, apart from the 2D mode width. As can be seen from Figure 4, the narrowest FWHM(2D) is confined to areas which show rather uniform Pos(G) and Pos(2D) and correspond to the flat regions on Cu surfaces (regardless the lattice orientation).

The XRD measurements show only small differences in crystalline quality between the oriented surfaces on all scales, which confirms that the obvious variations in strain and doping between graphene on the individual surface orientations are indeed driven by the particular alignment of graphene and Cu lattices. The distinct orientations of graphene with respect to underlying single crystalline Cu(100), Cu(110) and Cu(111) were also shown recently in ref. [32].

Finally, we turn our attention to the extremely narrow 2D band with the FWHM as low as 16 cm$^{-1}$ for graphene on Cu(100) (see Fig. 5). There are again clear differences between the tested Cu surfaces. The average FWHM(2D) increases from Cu(100) to Cu(110), further to Cu(111), with graphene on the Cu foil exhibiting the broadest 2D comparable to the commonly measured values with the average of ~ 31 cm$^{-1}$. The variations between the surfaces can be explained in part by different "uniformity" of graphene in terms of strain and doping at the scale of the laser spot (~ 1 μm), which can be observed as the statistical distributions in Fig. 3. The broader the distribution of data points, both horizontal (doping) and skewed (strain), the broader the 2D band, *cf.* the average FWHM(2D) in individual plots in Fig. 3. In other words, the FWHM(2D) reflects to some extent the microscopic spatial uniformity of graphene, which is projected onto the submicro- or nanoscale, i.e. below the laser spot size. This assumption is further corroborated by the correlation between FWHM(2D) and FWHM(G) (Figs 2f and S14, Supplementary Information). Figure S14 shows two basic trends connecting the G and 2D linewidths. The first trend is connected with the increasing spatial homogeneity under the laser spot, when both G and 2D bands are getting narrower. The second trend is visible when both FWHM(G) and FWHM(2D) reach their minima – for Cu(100). In that case, 2D band narrowing is accompanied by G band broadening caused by the lowered doping level in graphene down to $|E_F| \sim 0$ eV, when FWHM(G) $\sim 13$ cm$^{-1}$ and FWHM(2D) $\sim 16$ cm$^{-1}$. On the other hand, maximum doping level reached in our samples ($|E_F| \sim 350$ meV) is reflected in FWHM(G) drop down to $\sim 4.5$ cm$^{-1}$ and FWHM(2D) $\sim 20$ cm$^{-1}$ for graphene on Cu(100). It should be noted that in Fig. S14, all three main groups of points on Cu(100) with distinct doping levels are clearly separated (along the yellow arrow).

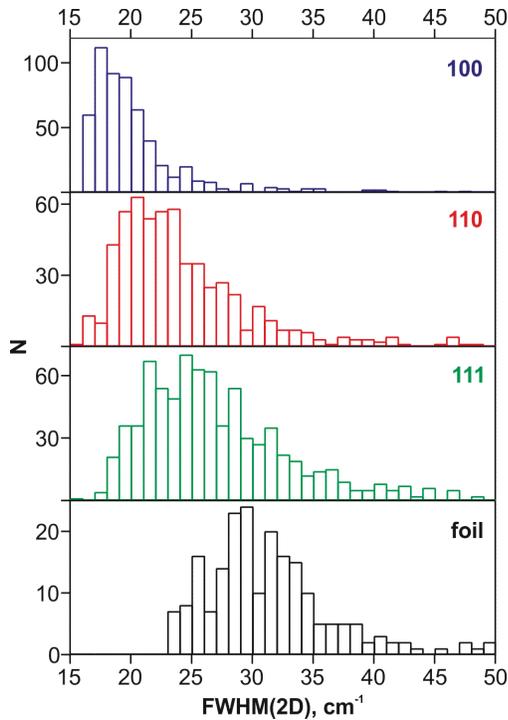

**Figure 5.** Histograms of FWHM(2D) accumulated from all measurements of as-grown graphene on a particular Cu surface. Excitation energy is 1.96 eV. The vertical axis shows the number of occurrences (N) of the particular FWHM(2D).

The exact reason for the low absolute values of FWHM(2D), especially on Cu(100), is not fully understood at the moment. However, one can expect to find the cause in the interaction between graphene and the metallic substrate. An exceptionally narrow 2D band was reported also in graphene on iridium [85]. A recent ARPES study of graphene on Cu foil showed a substantial reduction of electron-electron interaction in graphene by the metallic screening of the substrate and thus provided a more accurate analysis of electron-phonon interaction, which in turn showed a better convergence to local density approximation calculations [81]. We could therefore assume that the current theoretical approaches to the calculations and interpretations of the Raman spectra of graphene might actually reflect the situation on copper, at least in terms of the electron-phonon coupling strength. A simple formula for the FWHM(2D) calculation was provided by Basko, though only conical bands were taken into account [61]:

$$FWHM(2D) = 0.77 \frac{v_{ph}}{v_F} 8\gamma_{e-ph} \qquad (2)$$

Where $v_{ph}$ and $v_F$ are the phonon and Fermi group velocities, respectively, and $\gamma_{e-ph}$ is a broadening parameter due to electron-phonon scattering. The $v_{ph}/v_F$ is reflected in the slope of the 2D band dispersion with excitation energy, which is slightly higher for Cu(100) with the slope of 93.6 ± 0.6 cm$^{-1}$/eV in contrast to 88.7 ± 1.4 cm$^{-1}$/eV for Cu(111), caused by the higher doping level [78] (see Figure 6 left panel). However, such slopes are still very much in the range of those reported also for graphene on Si/SiO$_2$ [34]. The dependence of $\gamma_{e-ph}$ scattering term on the excitation energy was evaluated by Venezuela et al. [62]:

$$\gamma_{e-ph} = (18.88 E_L + 6.802 E_L^2)\ meV \qquad (3)$$

where $\gamma_{e-ph}$ in Eq. 2 corresponds to $\gamma_{e-ph}/4$ in Eq. 3. It should be noted that Basko [61] assumes the presence of two 2D components, each of them having the same FWHM calculated from Eq. 2, with their separation being a function of electron-hole asymmetry. The overall shape and width of the 2D peak stems from a complex interplay involving both electronic and phononic trigonal warping, which are opposite, i.e. they compensate each other to a varying extent (depending on $E_L$) [62], phonon self-energy corrections [86], polarization, instrumental broadening, etc. As a consequence, a quantitative comparison between the calculated and measured peak widths is usually avoided for doubly resonant scattering processes. In our case, a simple comparison of FWHM(2D) measured on graphene on Cu(111) and FWHM(2D) obtained from Eqs. 2 and 3 as a function of $E_L$ provides similar trends, i.e. the width increases with $E_L$ (Fig. 6 right panel). Nevertheless, the absolute values differ for the above mentioned reasons. On the other hand, in graphene on Cu(100) and (110), the FWHM(2D) increase with $E_L$ is only very small if present – the 2nd order polynomials can be fitted to the data with the minima at ~2.2 eV, but with adjusted $R^2$ of less than 0.85. (Due to the small significance of the fits, we do not plot them in Fig. 6, in contrast to Cu(111) with adj. $R^2 > 0.99$). The different scaling of FWHM(2D) with $E_L$ might indicate a weaker dependence of electron-phonon coupling in graphene on Cu(100) and Cu(110) and/or significant variations in the shapes of electronic and phononic bands resulting from the distinct orientation-dependent interaction strengths between graphene and copper. Both these assumptions can be justified by the data presented here as well as in the literature [24, 30, 45, 81, 82], and their separation requires further detailed experiments together with complex models embracing a wider range of parameters.

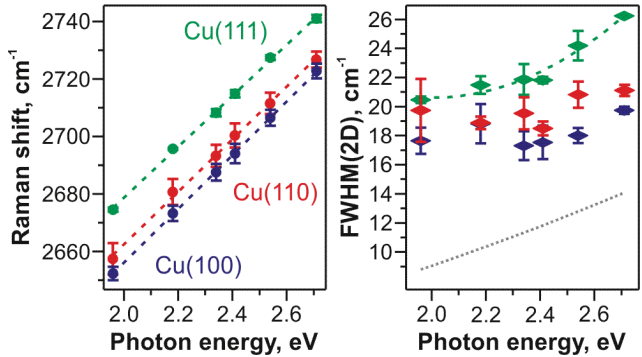

**Figure 6.** Dispersion of 2D band position (left) and FWHM (right) with laser excitation energy for Cu(100), Cu(110) and Cu(111) plotted with blue, red and green, respectively. Dashed lines are linear and 2nd order polynomial weighted least-squares fits to experimental points in the left and right panels, resp. Error bars represent standard deviation from 4 measurements. The grey dotted line in the right panel is a calculation according to Eq. 2, using the experimental 2D band dispersion for graphene on Cu(111) and $\gamma_{e-ph}$ from Eq. 3.

## 4. Conclusions

We have shown that graphene grown by CVD on copper is highly susceptible to the lattice orientation of the substrate, which influences the strain and doping in graphene – both locally and generally. As expected, the growth on Cu single crystals, regardless of the orientation, provides more homogeneous graphene as compared to the commonly used polycrystalline Cu foil. Graphene on Cu(111) is mainly uniformly compressed with ~0.3% of biaxial strain and n-doped with Fermi level shift of ~250 meV, though with minor occurrences of relaxed and wrinkled regions. Graphene on Cu(110) is compressed with a large range of values between 0.05 and 0.3% of biaxial strain and essentially undoped. Graphene on Cu(100) is slightly and uniformly compressed (0.1%) and occurs in three states with distinct doping levels, where the majority (~60%) of the graphene is undoped and the other two parts, equal in appearance, show n-doping with Fermi level shifts of ~ 250 and 350 meV, respectively. The data gathered for graphene on Cu(100) and Cu(111) are in agreement with the few previous studies by ARPES or STM and show that Raman spectroscopy provides quantitative information when done on data sets large enough.

An important result, which can have serious implications for the understanding of the origin of the double resonance 2D (or G') band, is the extremely narrow width of this band on all three single crystals, especially on Cu(100). Nevertheless, more detailed studies of the interaction between graphene and the metallic substrate are necessary to fully describe the observed phenomenon.

**Acknowledgments**

Financial support was provided by the Grant Agency of the Czech Republic (Contract No. P204/10/1677), FP7-Energy-2010-FET project Molesol (Contract No. 256617), and Ministry of Education, Youth and Sports ERC-CZ project (Contract No. LL1301).

**Appendix A. Supplementary Data**: Supporting data associated with this article (Laue images, reciprocal and real space XRD maps of Cu surfaces; additional correlations of selected parameters from Raman spectra of graphene on Cu single crystals, AFM images), can be found in the online version, at http://dx.doi.org/10.1016/*************.

.

Electronic Supplementary Information

# Interaction between graphene and copper substrate: The role of lattice orientation

*Otakar Frank[1,*], Jana Vejpravova[2], Vaclav Holy[3], Ladislav Kavan[1], and Martin Kalbac[1]*

[1] J. Heyrovsky Institute of Physical Chemistry of the AS CR, v.v.i., Dolejskova 2155/3, CZ-182 23 Prague 8, Czech Republic

[2] Institute of Physics of the AS CR, v.v.i., Na Slovance 1999/2, CZ-182 21 Prague 8, Czech Republic

[3] Faculty of Mathematics and Physics, Charles University in Prague, Ke Karlovu 5, CZ-121 16 Prague 2, Czech Republic

* Corresponding author: otakar.frank@jh-inst.cas.cz

**Table S1.** Samples of graphene grown on copper substrates for Raman mapping with laser excitation of 1.96 eV. All the samples in one batch were grown in the same time and measured in the order as indicated. The samples in queue for characterization were stored under vacuum. The order of samples for measurement was varied amongst the batches to check if there is any influence of the waiting time on the results of Raman characterization. No such effects were observed. The numbers below sample orientation indicate the size of Raman maps as X x Y (lateral step size of 1 µm in both directions). Total measurement time for one spectrum was ~ 260 s (two spectral windows of 120 s each plus grating movement).

| Batch # | Measured as | | | |
|---|---|---|---|---|
| | 1st | 2nd | 3rd | 4th |
| 1 | Cu(100) 16x16 | Cu(110) 17x16 | Cu(111) 18x17 | -- |
| 2 | Cu(111) 14x14 | Cu(110) 12x11 | Cu(100) 14x14 | Cu foil 15x14 |
| 3 | Cu(110) 14x14 | Cu(100) 11x11 | Cu(111) 14x14 | -- |

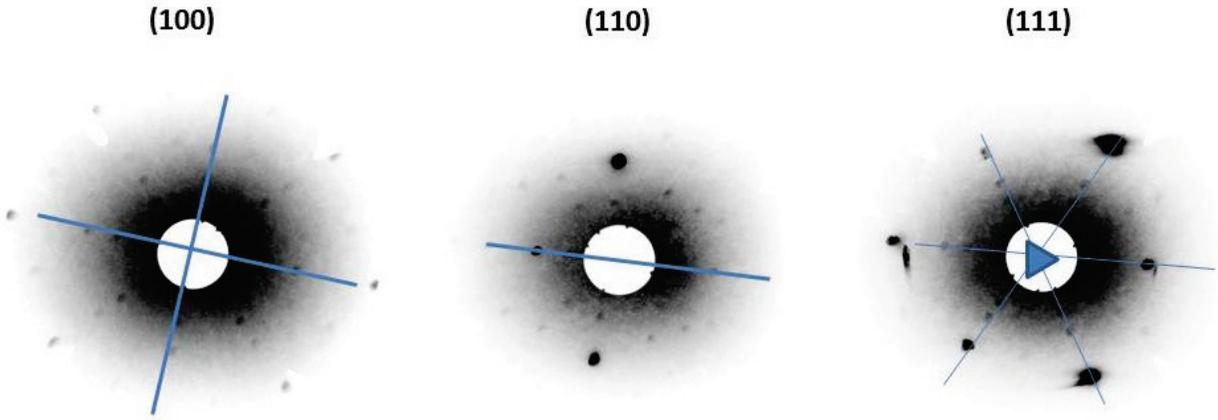

**Figure S1.** Typical Laue images of the copper single-crystals recorded with the incident beam perpendicular to the (100), (110) and (111) planes. The patterns clearly reflect the expected minimum symmetry of the corresponding Laue class: *m-3m*.

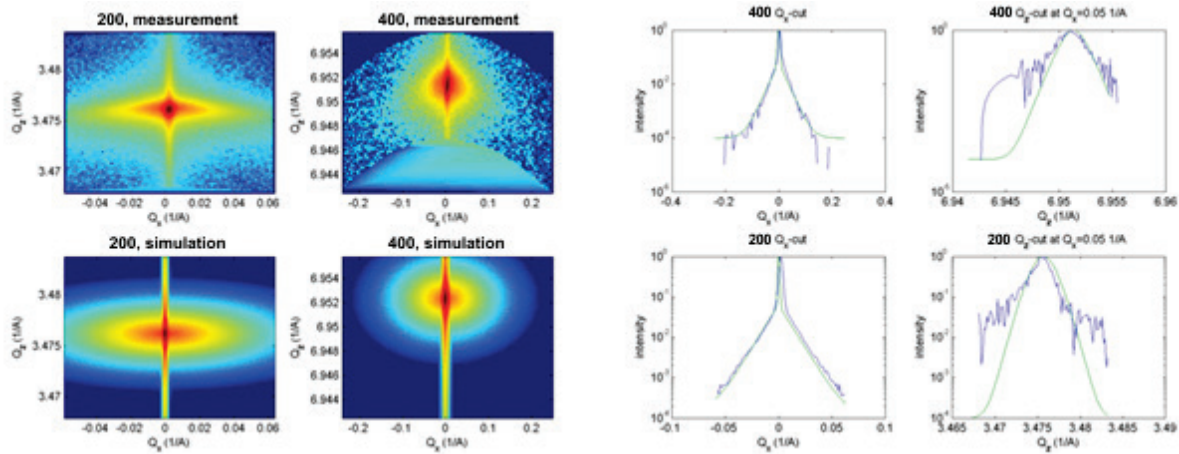

**Figure S2.** Reciprocal maps of the copper single-crystals for the (100) plane. The left panel corresponds to the experimental data (top) and simulation (bottom) for the (200) and (400) diffractions, the right panel shows the corresponding fits of a single scan along the $Q_x$ or $Q_z$ direction in reciprocal space.

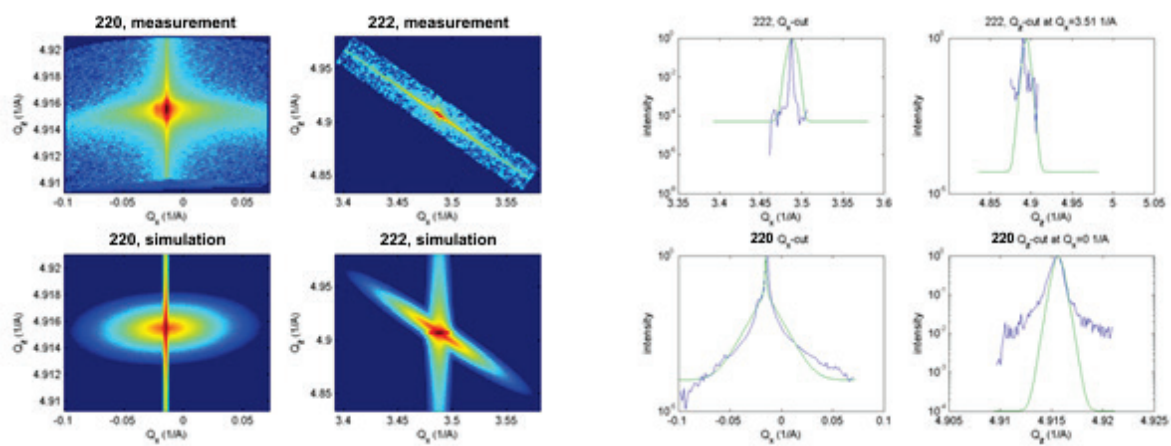

**Figure S3.** Reciprocal maps of the copper single-crystals for the (110) plane. The left panel corresponds to the experimental data (top) and simulation (bottom) for the (220) and (222) diffractions, the right panel shows the corresponding fits of a single scan along the $Q_x$ or $Q_z$ direction in reciprocal space.

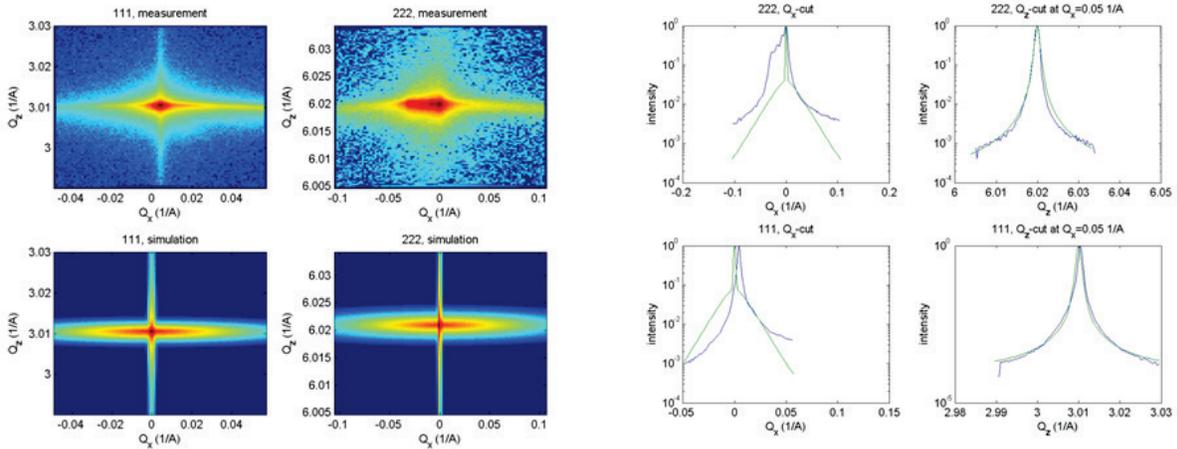

**Figure S4.** Reciprocal maps of the copper single-crystals for the (111) plane. The left panel corresponds to the experimental data (top) and simulation (bottom) for the (111) and (222) diffractions, the right panel shows the corresponding fits of a single scan along the $Q_x$ or $Q_z$ direction in reciprocal space.

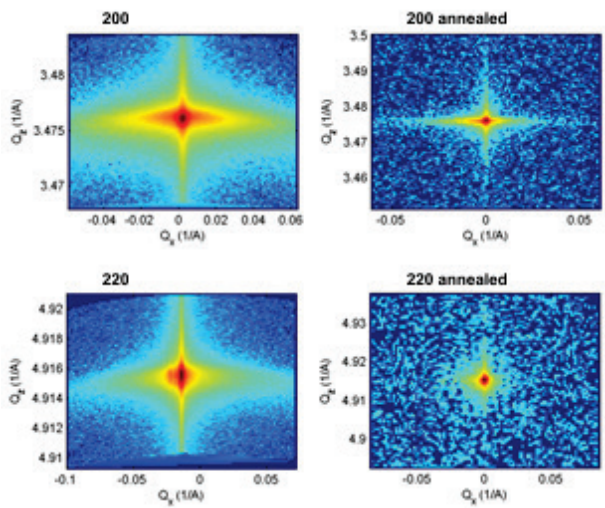

**Figure S5.** Reciprocal maps of the copper single-crystals for the (100) and (110) faces recorded on the samples before and after annealing, respectively.

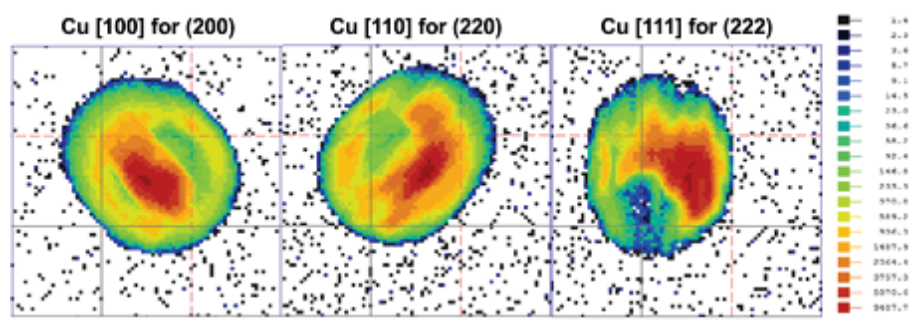

**Figure S6.** Real space maps of the copper single-crystals recorded perpendicular to the (100), (110) and (111) planes. The lateral dimensions of the maps is 10x10 mm$^2$.

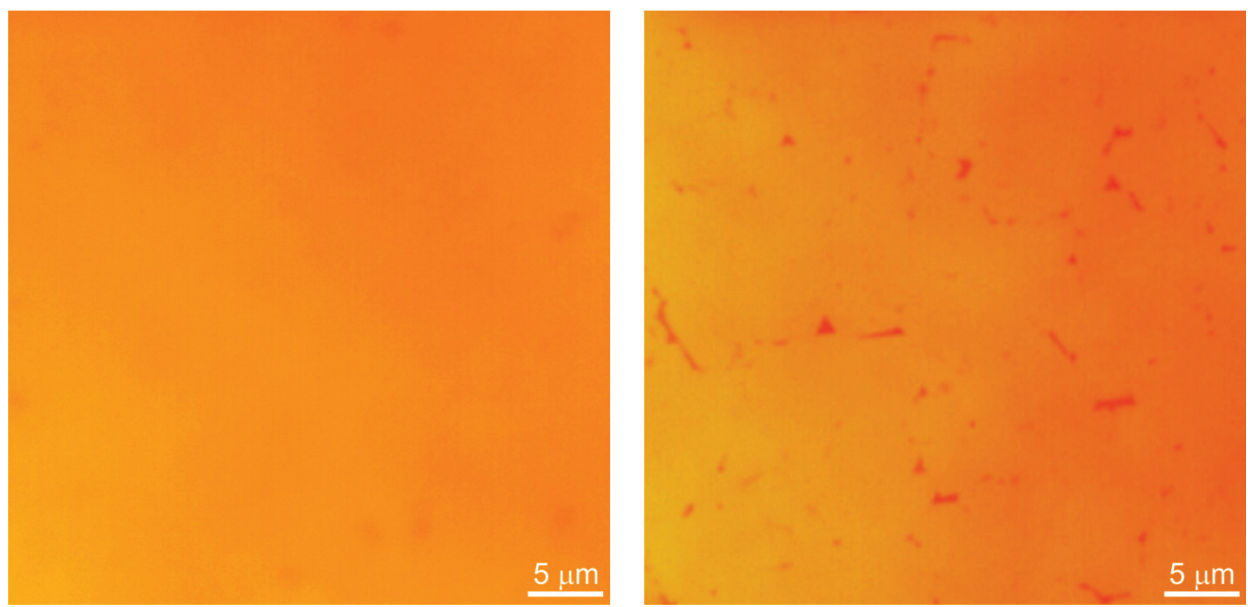

**Figure S7**. Optical microphotographs of Cu(111) surface with freshly grown graphene (left) and after 1 week of storing at ambient conditions (different area).

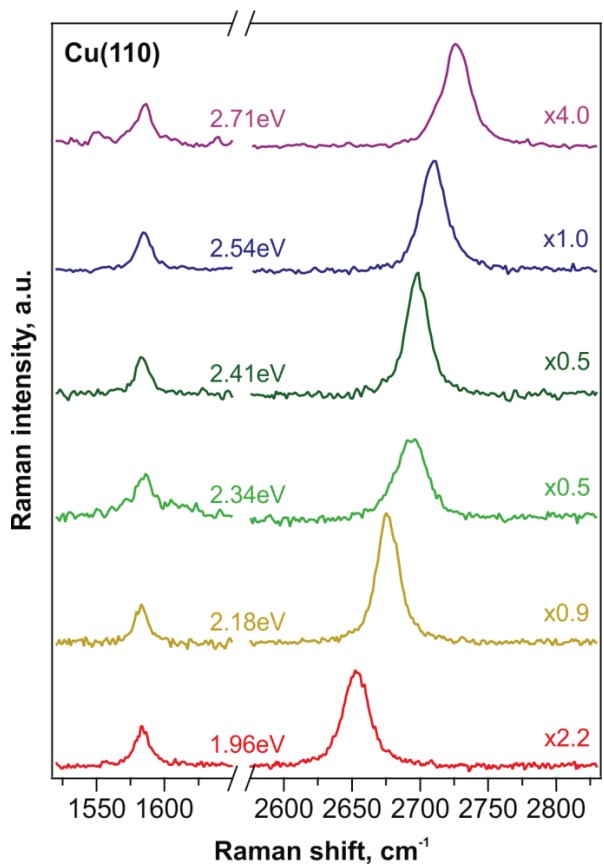

**Figure S8.** Raman spectra of graphene grown on Cu(110) single crystal. Excitation energies are given for each spectrum, the spectra are normalized to the G band intensity for each Cu surface, the corresponding zoom factors are labeled in the plots.

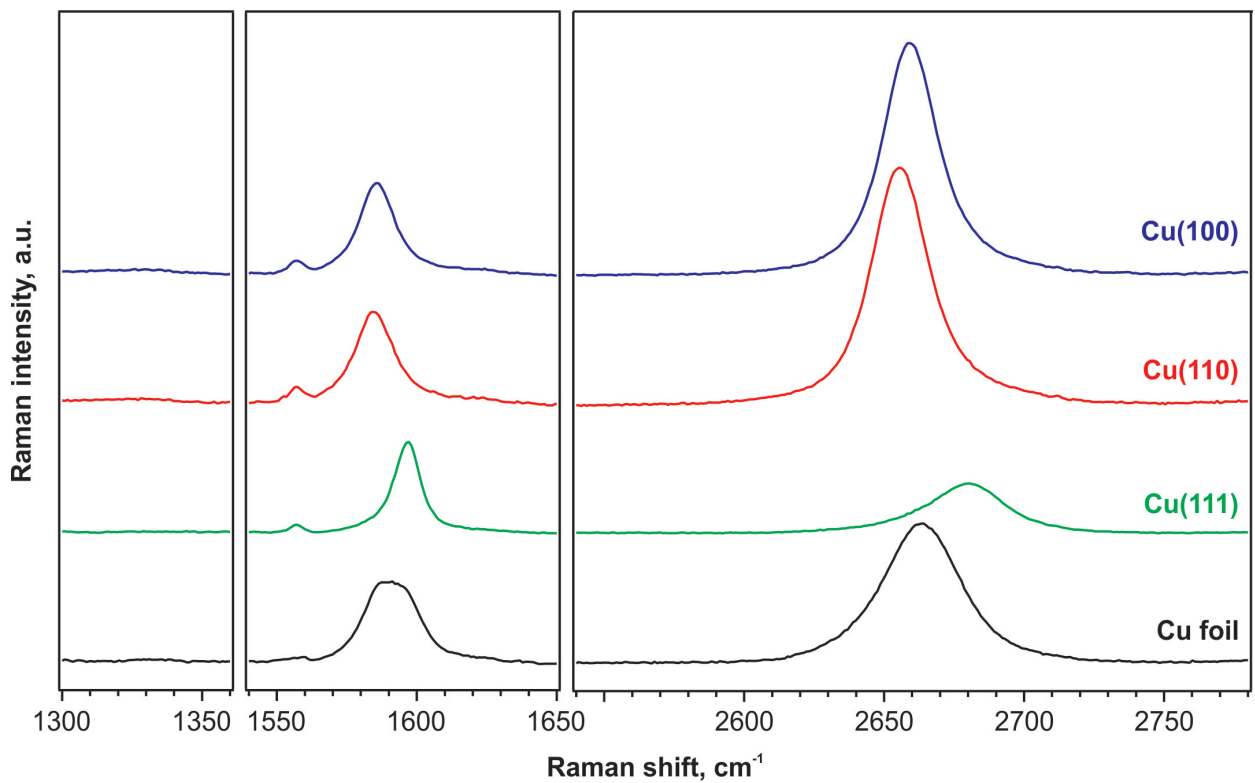

**Figure S9.** Average Raman spectra obtained by averaging all spectra of graphene acquired during mapping of batch #2 on Cu(100) (blue), Cu(110) (red), Cu(111) (green), and Cu foil (black). Laser excitation energy is 1.96 eV. Note that the peaks are much broader than what would correspond to single spectra because of the span of strain (and doping) levels across the sampled areas. The peak at ~1555 cm$^{-1}$ corresponds to the vibration of molecular oxygen.

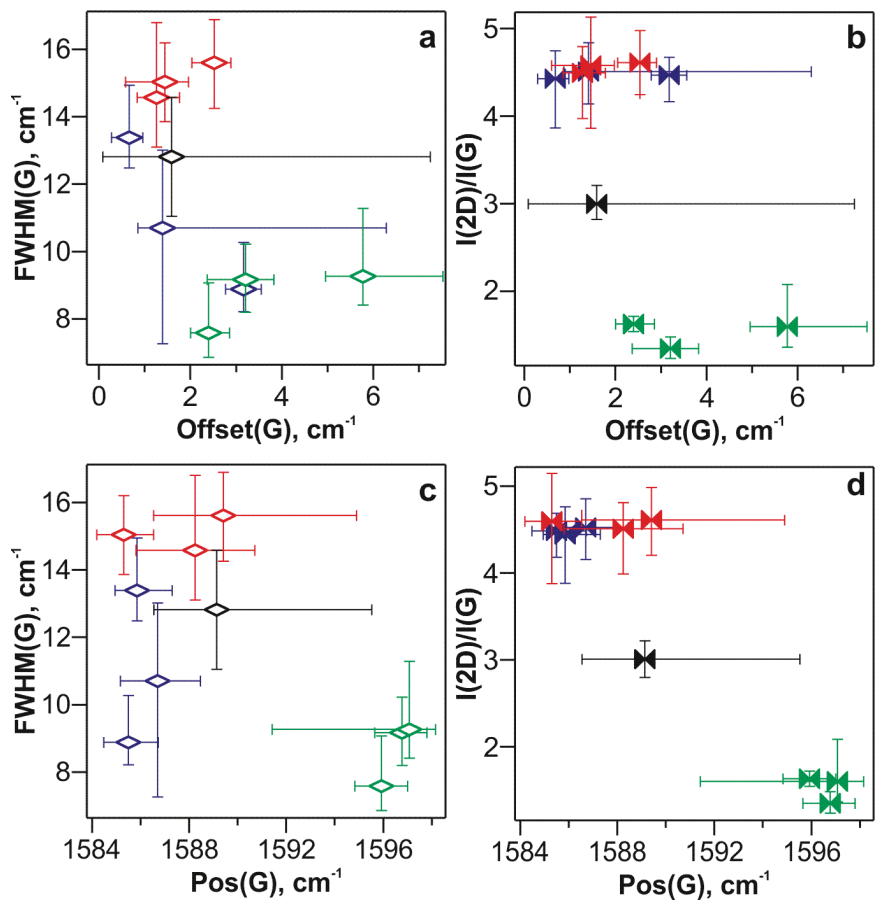

**Figure S10.** Correlations between monitored parameters of the G peaks fitted by Lorentzian lineshapes in Raman spectra of as-grown graphene on Cu(100) (blue), Cu(110) (red), Cu(111) (green) and Cu foil (black). To show the difference between Pos(G) and Offset(G), panels a and b are exact copies of Fig. 2e and b, respectively, with the panels c and d showing the same data, but the G band frequencies displayed as the "absolute" Raman shifts (i.e. Pos(G)) as opposed to the Offset(G), which is the relative difference from the shift caused purely by strain (for further explanation, see main text).

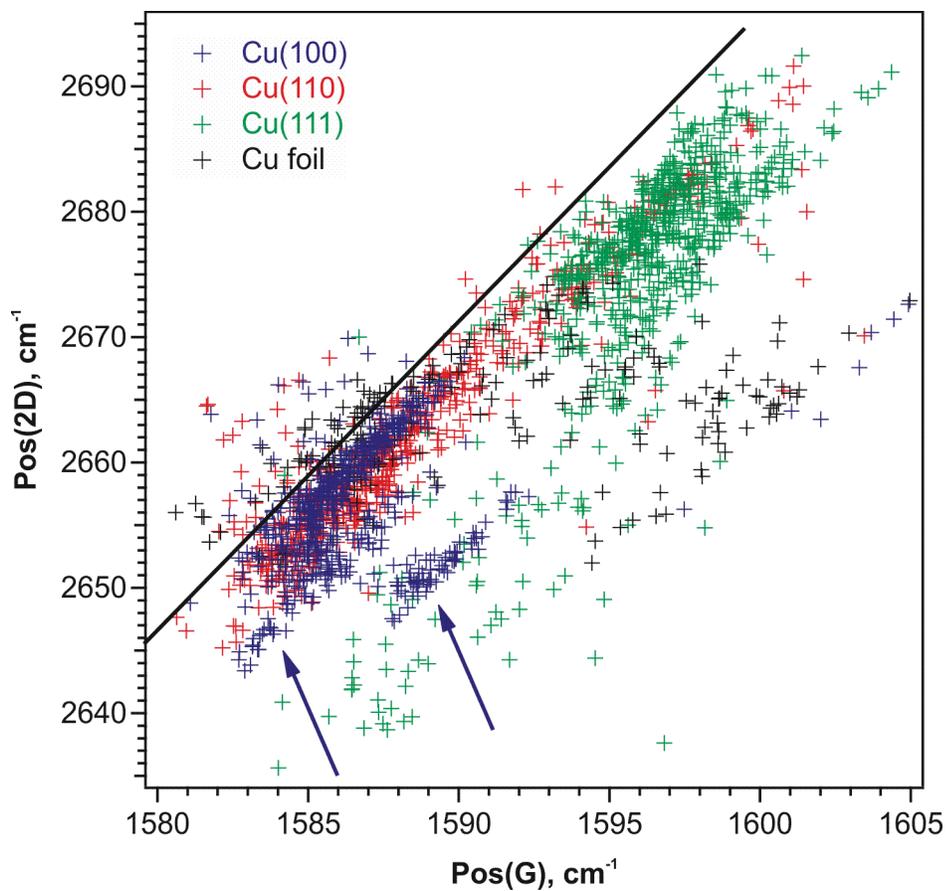

**Figure S11.** Correlation between Pos(2D) and Pos(G) for all measured data points on the studied Cu surfaces differentiated by colors. Laser excitation energy is 1.96 eV. The black line indicates pure strain variation, with the slope of 2.5. Blue arrows label two data sets with doping levels distinct from the majority of points on Cu(100).

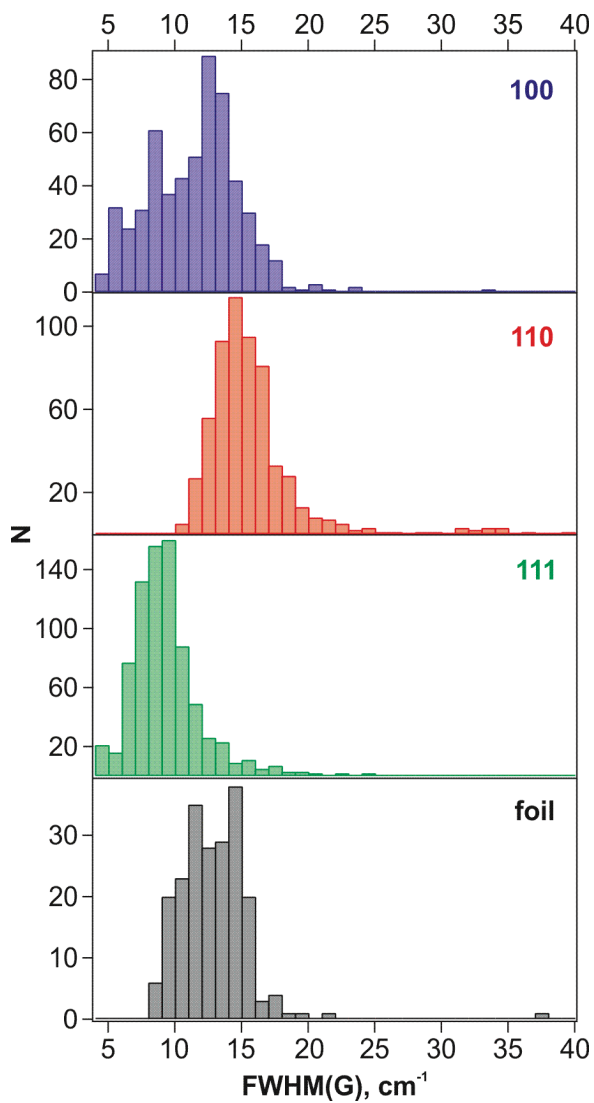

**Figure S12.** Histograms of FWHM(G) accumulated from all measurements of as-grown graphene on a particular Cu surface. Laser excitation energy is 1.96 eV. The vertical axis shows the number of occurrences (N) of the particular FWHM(G).

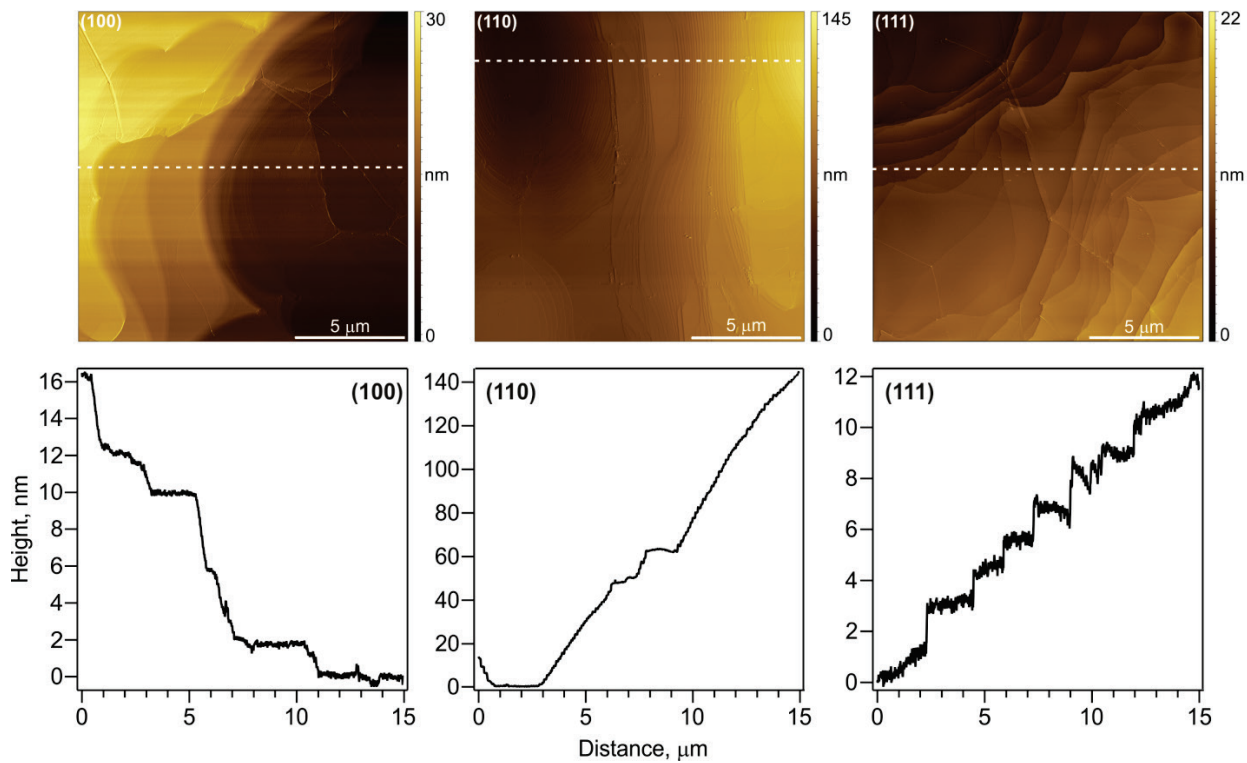

**Figure S13.** Top: AFM images (height sensor) of graphene on Cu(100), Cu(110) and Cu(111) single crystals. Bottom: height profiles extracted along the dashed white lines from the respective AFM images. Note the different height scales (both in images and profiles) when comparing between the individual Cu orientations.

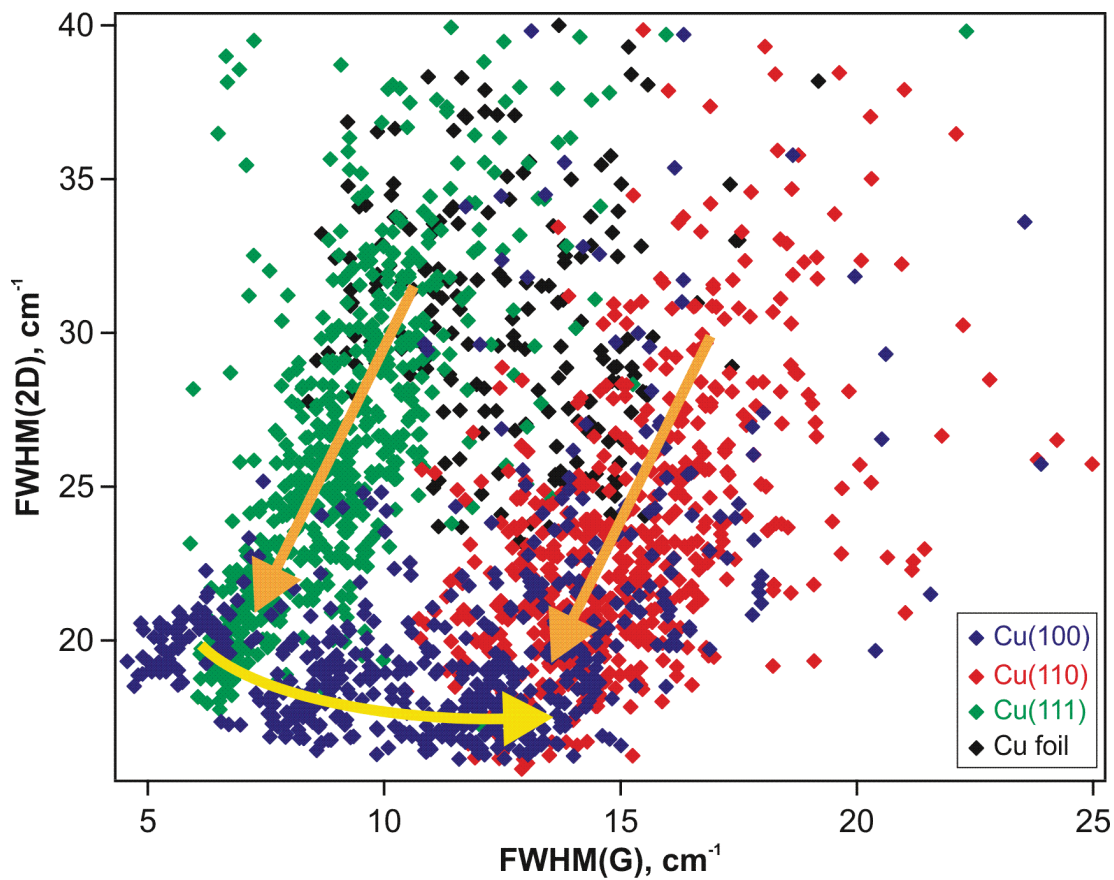

**Figure S14.** Correlation between FWHM(2D) and FWHM(G) for all measured data points on the studied Cu surfaces differentiated by colors. Laser excitation energy is 1.96 eV. The orange and yellow arrows depict the two main trends in 2D band narrowing: increasing the spatial homogeneity (orange) accompanied by G band narrowing and reducing the doping level (yellow) accompanied by G band broadening.